\documentstyle[aps,pre,psfig]{revtex}

\def\meanm#1{\langle #1\rangle}
\begin{document}
\draft
\twocolumn[
\null
\begin{center}
\preprint{HEP/123-qed}
{\large\bf Viscous Flow and Jump Dynamics in Molecular 
Supercooled Liquids:\par II Rotations \par}
\vskip 1.5em
Cristiano De Michele${}^{1}$ and Dino Leporini${}^{1,2}$ \cite{byline}\par
{\it
${}^{1}$ Dipartimento di Fisica, Universit\`a di Pisa, V.Buonarroti, 2 
I-56100 Pisa, Italy \par}
{\it
${}^{2}$ Istituto Nazionale di Fisica della Materia, Unit\`a di Pisa \par}
{( Received \today ) \par}
\vskip 0.9em
\begin{minipage}{14.1cm}
\small
The rotational dynamics of a 
supercooled model liquid of rigid A-B dumbbells interacting via a L-J 
potential is investigated along one single isobar. The
orientation correlation functions exhibit a two-step character which 
evidences molecular trapping. Trapping is also apparent in a suitable 
angular-velocity correlation functions. The rotational correlation 
$\tau_{1}$ is found to scale over more than three orders of magnitude as 
$(T-T_{c})^{-\gamma}$ with $\gamma = 1.47 \pm 0.01$ and $T_{c}$ the 
MCT critical temperature. Differently, $\tau_{l}$ with $l=2-4$ and 
the rotational diffusion coefficient $D_{r}$ manifest deviations.
For $0.7 < T < 2$ good agreement with the diffusion model is found, 
i.e. $\tau_{l} \cong 1/ l(l+1) D_{r}$. For lower temperatures the 
agreement becomes poorer and the results are also only partially 
accounted for by the jump-rotation model.  The angular Van Hove function 
evidences that in this region a meaningful fraction of the sample 
reorientates by jumps of about $180^{\circ}$. The distribution 
of the waiting-times in the angular sites cuts exponentially at long
times. At lower temperatures it decays at short times as 
$t^{\xi-1}$ with  $\xi = 0.34 \pm 0.04$ at $T=0.5$ in analogy with 
the translational case. The breakdown of the 
Debye-Stokes-Einstein is observed at lower temperatures where
the rotational correlation times diverge more weakly than the 
viscosity.
\end{minipage}
\end{center}
\par
]

\section{INTRODUCTION}
\label{sec:intro}

The relaxation and the transport properties of molecular liquids 
depend on both their translational and rotational motion. Since their
mutual interplay cannot be neglected both dynamical aspects must be 
jointly considered. If the liquids are supercooled or supercompressed the 
overwhelming difficulties to molecular rearrangement are expected to 
enhance the role of the rotational degrees of freedom and, more 
particularly, the rotational-translational coupling as noticed by 
experiments \cite{review,ediger,sillescu1,tork3,euro}, 
theory \cite{schillingrot,gotzerot,dieze} and numerical 
\cite{bagchi,kammererrot1,kammererrot2,kammerertrasl} studies and 
discussed in a recent topical meeting \cite{noi}.

Molecular-dynamics numerical ( MD ) studies provided
considerable insight in supercooled liquids during the 
last years \cite{kob}. However, the question of rotational 
dynamics seems to have been partially 
overlooked since most studies dealt with atomic systems where 
rotational dynamics is missing. Notable exceptions addressed the 
issue in model systems of disordered dipolar lattice \cite{bagchi}, 
biatomic molecules \cite{kammererrot1,kammererrot2} and well 
studied glassformers, e.g. CKN  \cite{signorini}, OTP \cite{lewis,kud} 
and methanol \cite{sind} . The case of supercooled water was 
investigated in detail \cite{sciortino}. Studies of plastic crystals 
and orientational glasses ( i.e. no allowed translation ) are also 
known \cite{renner,lee}. 

We have recently presented numerical results on the the translational 
motion of a supercooled molecular model liquid 
\cite{demichel2} ( hereafter referred to as I ). The present paper 
wishes to complement I by extending the analysis 
to rotational degrees of freedom. As in I one issue is the
detection and characterization of jump dynamics. It is found that 
rotational jumps are fairly more frequent than translational ones in 
the present system. This makes it easier their study. The occurrence of 
jumps poses the question of the coupling of the molecular reorientation 
with the shear viscous flow. This is the second issue addressed in 
the paper. Jump dynamics may take place in the absence of any shear flow.
Nonetheless, shear motion may favour jumps over energy barriers 
\cite{eyring}. The question is of relevance in that the experimental 
situation is rather controversial. For macroscopic bodies hydrodynamics 
predicts that the reorientation is strongly coupled 
to the viscosity $ \eta $ according to the 
Debye-Stokes-Einstein law ( DSE ),
$ \tau, D_{r}^{-1} \propto \eta $, where $ \tau$ and  $D_{r}$
are the rotational correlation time and diffusion coefficient, 
respectively \cite{lamb,favro,hu}.
DSE is quite robust. In fact, the coupling of the reorientation 
to the viscosity is usually found even at a molecular level 
if the viscosity is smaller of about $1-10 \; Poise$. 
At higher values DSE overestimates the correlation times of 
tracers in supercooled liquids according to time-resolved 
fluorescence \cite{ye,tork2} and Electron Spin Resonance 
( ESR ) studies \cite{euro,vigo,macromol}. On the other hand, 
photobleaching \cite{ediger} and NMR \cite{sillescu1}
studies found only small deviations from DSE even close to $T_{g}$. 
Interestingly, according to ESR studies 
in the region where tracer reorientation decouples 
by the viscosity, an ESR study evidenced that it occurs 
by jump motion \cite{jumpesr,ivanov}. 

The paper is organized as follows. In section \ref{sec:moddet} 
details are given on the model and the simulations. In Sec. 
\ref{sec:results} and \ref{sec:conclusion} the results are discussed 
and the conclusions are summarized, respectively. 

\section{MODEL AND DETAILS OF SIMULATION}
\label{sec:moddet}

The system under study is a model molecular liquid of rigid 
dumbbells \cite{kammererrot1,kammererrot2,kammerertrasl}.
The atoms A and B of each molecule have mass $m$ and are spaced 
by $d$. Atoms on different molecules interact via the 
Lennard-Jones potential:

\begin{equation}
V_{\alpha\beta}(r) = 4\epsilon_{\alpha\beta} \left[ 
{(\sigma_{\alpha\beta}/r)}^{12} 
- {(\sigma_{\alpha\beta}/r)}^6\right],\quad \alpha,\beta \in \{A,B\}
\end{equation}

\noindent The potential was cutoff and shifted at 
$r_{cutoff}=2.49\sigma_{AA}$. Henceforth, reduced units will 
be used. Lenghts are in units of $\sigma_{AA}$, energies in
units of $\epsilon_{AA}$ and masses in units of $m$. The time 
unit is  $\left(\frac{m \sigma_{AA}^2}{\epsilon_{AA}}\right)^{1/2}$,
 corresponding to about $2 ps$ for the Argon atom.
The pressure $P$, temperature $T$ and shear viscosity 
$\eta$ are in units of $\epsilon_{AA}/ \sigma_{AA}^{3}$ ,
$\epsilon_{AA} /k_{B}$ and $\sqrt{ m \epsilon_{AA}}/ \sigma_{AA}^{2}$, 
respectively.

The model parameters in reduced units are: 
$ \sigma_{AA}=\sigma_{AB} = 1.0$, $\sigma_{BB}= 0.95$, 
$\epsilon_{AA}= \epsilon_{AB}=1.0$, $\epsilon_{BB}=0.95$, $d = 0.5$, 
$m_A = m_B = m = 1.0$. 
The  $\sigma_{AA}$ and $\sigma_{BB}$ values were chosen 
to avoid crystallization. The sample has 
$N = N_{at}/2 = 1000$ molecules which are accommodated in a 
cubic box with periodic boundary conditions. Further details on the 
simulations may be found in I.

We examined the isobar at $P = 1.5$ by equilibrating
the sample under isothermal-isobaric conditions and then 
collecting the data by a production run in microcanonical 
conditions. The temperatures we investigated 
are $T= 6, 5, 3, 2, 1.4, 1.1, 0.85, 0.70, 0.632, 0.588, 
0.549, 0.52, 0.5$.

\section{RESULTS AND DISCUSSION}
\label{sec:results}

This section will discuss the results of the study. We 
characterize the correlation losses of the system by investigating 
several rotational correlation functions. Then, the related correlation 
times and transport coefficients will be presented. The 
presence of rotational jumps will be evidenced and their waiting time 
distribution will be discussed. Finally, the decoupling of the 
transport and the relaxation from the viscous flow will be presented.   

\subsection{Correlation Functions}
\label{sec:corrfunc}

The rotational correlation loss is conveniently presented  
by suitable correlation functions. We study  the dynamics of both the 
orientation and the angular velocity of the dumb-bell.

\subsubsection{Orientation}
\label{sec:orient}

The rotational correlation functions are defined:

\begin{equation}
C_l (t) = \frac{1}{N} \sum_{i=1}^{N} \left\langle P_l({\bf u}_i(t)
\cdot {\bf u}_i(0))\right\rangle 
\end{equation}

\noindent ${\bf u}_i(t)$ is the unit vector parallel to 
the axis of the molecule $i$ at time $t$ and $P_l(x)$
the Legendre polynomial of order $l$. It is worth noting 
that $C_1$ and $C_2$ are accessible to several experimental techniques, 
e.g. dielectric spectroscopy, NMR, ESR, light and neutron scattering.

\begin{figure}
\psfig{file=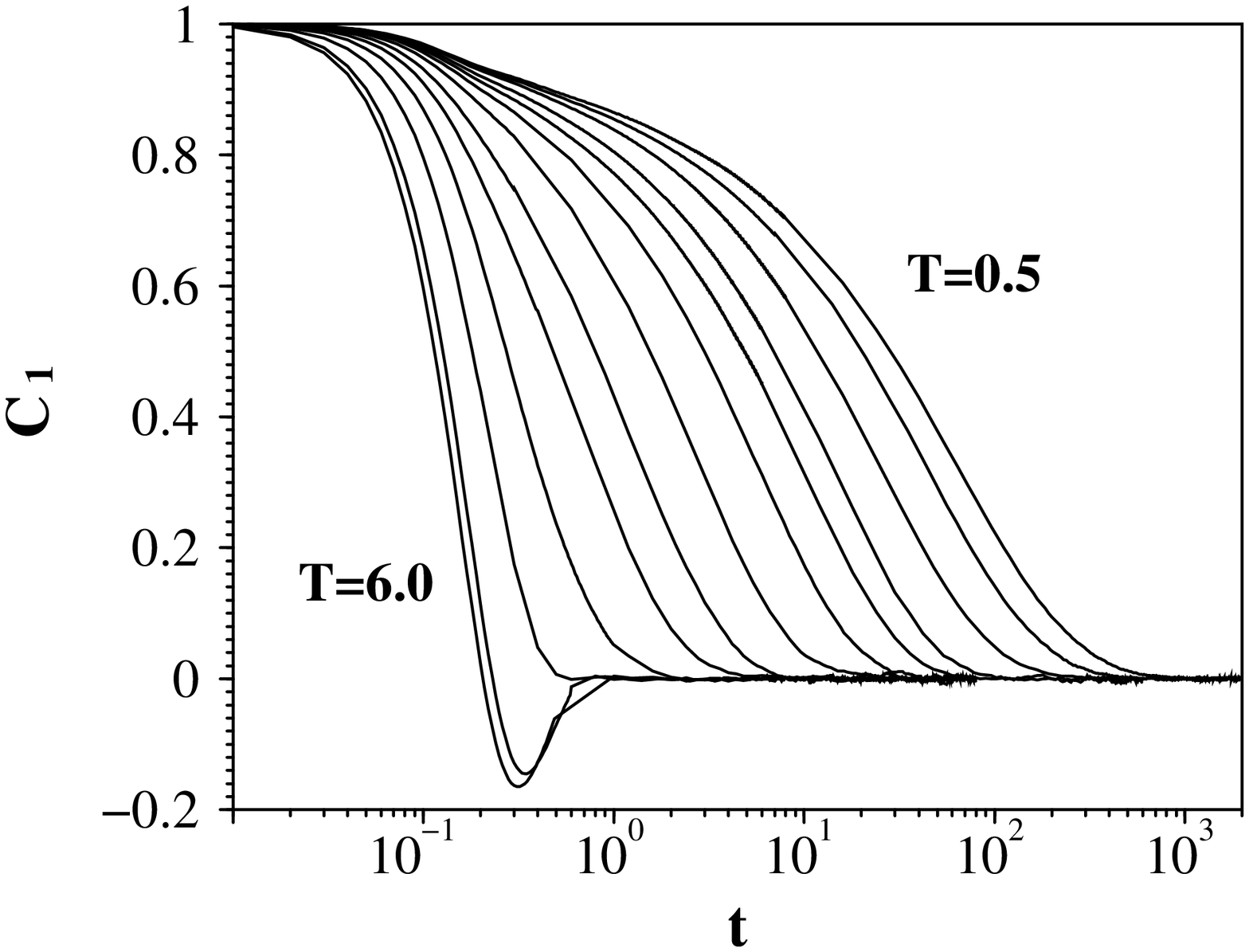,width=8cm,height=6.5cm}
\psfig{file=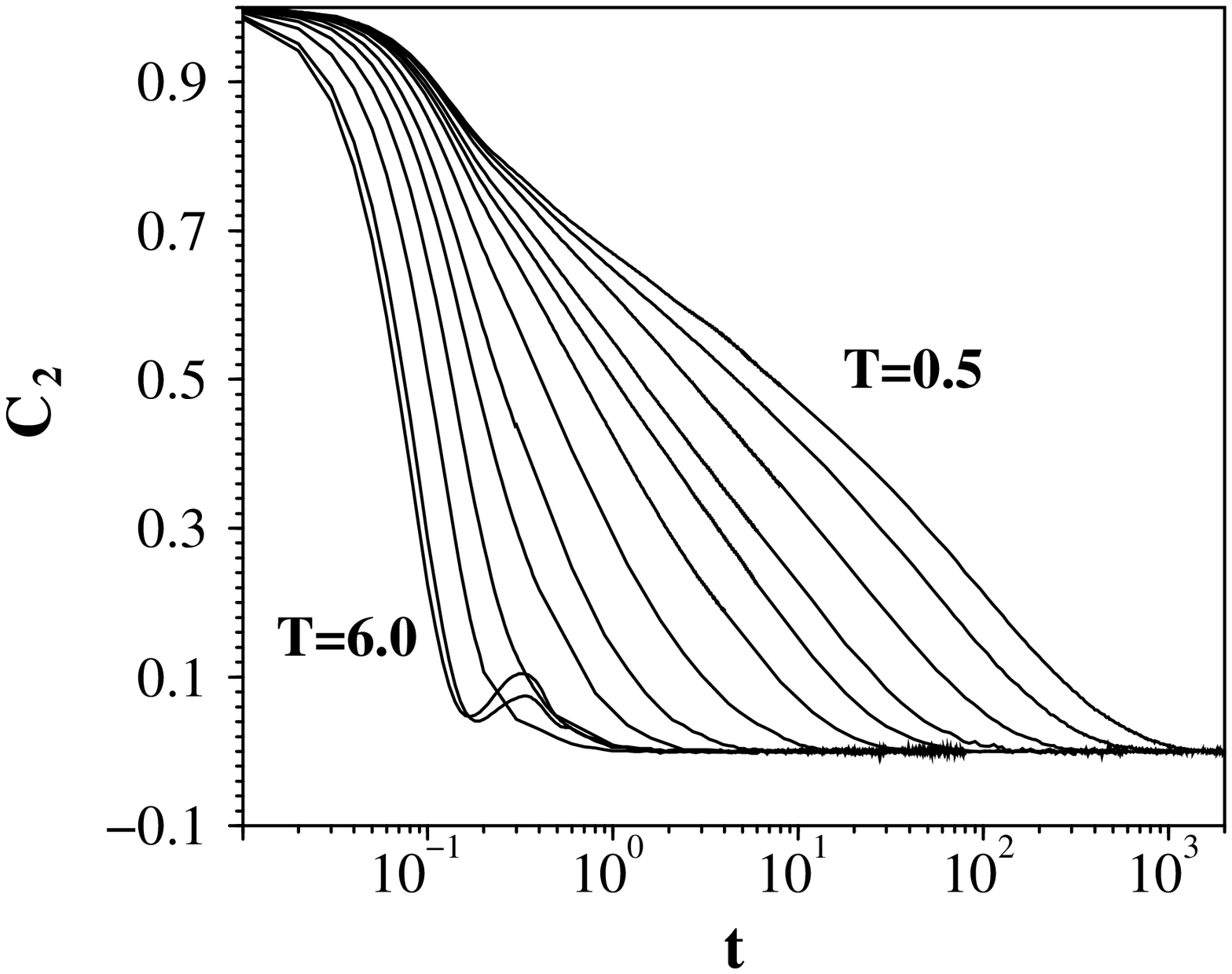,width=8cm,height=6.5cm}
\caption{Correlation functions $C_1$ and $C_2$ 
plotted for all the temperatures investigated.}
\label{figc12}
\end{figure}

Fig. \ref{figc12} shows $C_1$ and $C_2$. At high temperature and short 
times damped oscillations are present. They are typical features 
of free rotators in gas-like systems \cite{renner,hansen}.
At lower temperatures and intemediate times $C_1$ and $C_2$ exhibit a wide 
plateau which evidences the increased angular trapping. At longer times 
the decay is fairly well described by the stretched exponential 
$\sim exp[-(t/\tau)^{\beta}]$ with $\tau = 62.9$, $\beta = 0.70$ for 
$l=1$ and $\tau = 96.1$, $\beta = 0.60$ for $l=2$  at $T=0.5$.
Even if $C_1$ and $C_2$ are quite similar two differences must be noted. 
First the plateau is lower at $l=2$ than at $l=1$. Second, at lower 
temperatures $C_2$ vanishes at longer times than $C_1$. The first 
feature is understood by noting 
that the oscillatory character of $P_l({\bf u}_i(t) \cdot {\bf u}_i(0))$
with respect to the angle between ${\bf u}_i(t)$ and ${\bf u}_i(0)$ 
increases with $l$. Then, by increasing $l$ even random angular changes 
with small amplitude occurring at short times affect the decay of
$C_l$. The second feature is due to the fact that, as it will be 
shown later, molecules undergo frequent $180^{\circ}$ flips at lower 
temperatures. Due to the nearly head-tail symmetry, the flips reverse 
the sign of $P_l({\bf u}_i(t) \cdot {\bf u}_i(0))$ if $l$ is odd 
whereas no change takes place if $l$ is even. Then, they mainly affect
the decay of $C_l$ with odd $l$ values.

\begin{figure}
\psfig{file=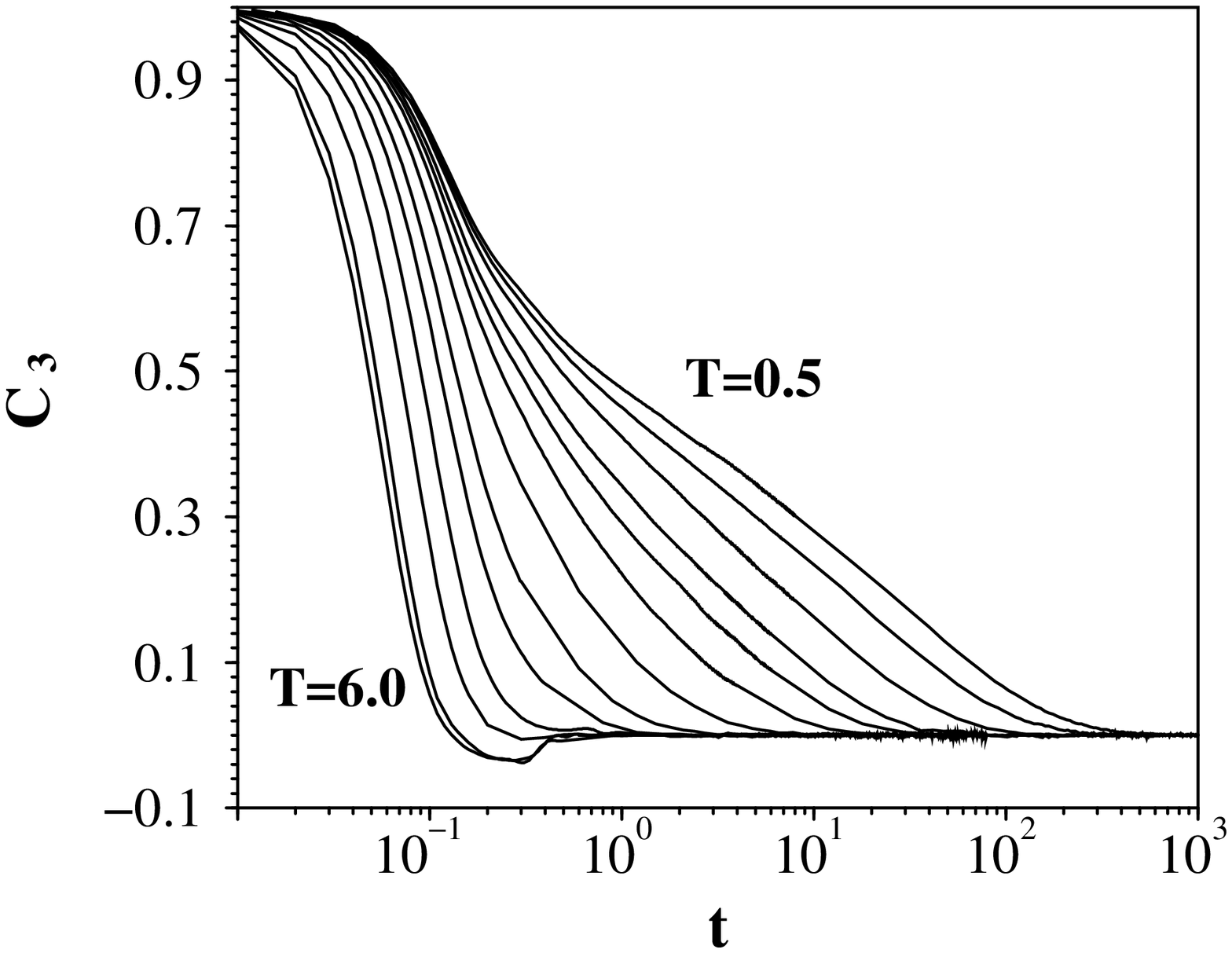,width=8cm,height=6.5cm}
\psfig{file=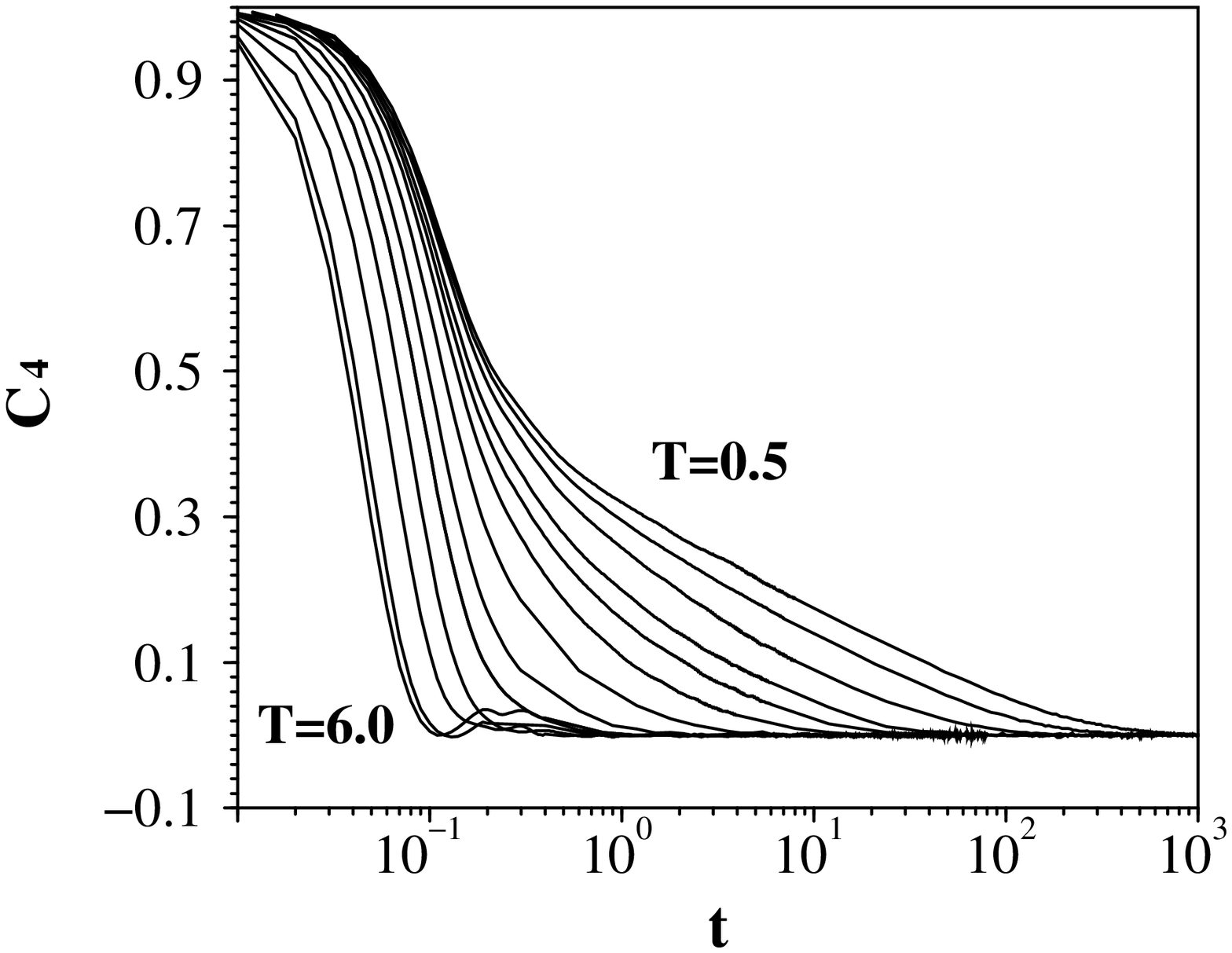,width=8cm,height=6.5cm}
\caption{Correlation functions $C_3$ and $C_4$ for all 
the temperatures investigated.}
\label{figc34}
\end{figure}

Fig. \ref{figc34} plots the functions $C_3$ and $C_4$. The 
discussion is similar to the case $l=1,2$. We note that, as expected, 
the plateau decreasess by increasing $l$. On the basis of the 
above discussion $C_3$ should vanish before $C_4$. However, on 
increasing $l$, the effect of the head-tail symmetry is partially 
masked by the increased oscillatory character of the Legendre 
polynomials, which yields a larger sensitivity to small-angle 
reorientations. Similarly to $C_{1,2}$ at longer times 
the decay is fairly well described by the stretched exponential 
$\sim exp[-(t/\tau)^{\beta}]$ with $\tau = 32.4$, $\beta = 0.60$ for 
$l=3$ and $\tau = 27.9$, $\beta = 0.47$ for $l=4$ at $T=0.5$. We notice 
that the stretching parameter decreases with increasing $l$.

\begin{figure}
\psfig{file=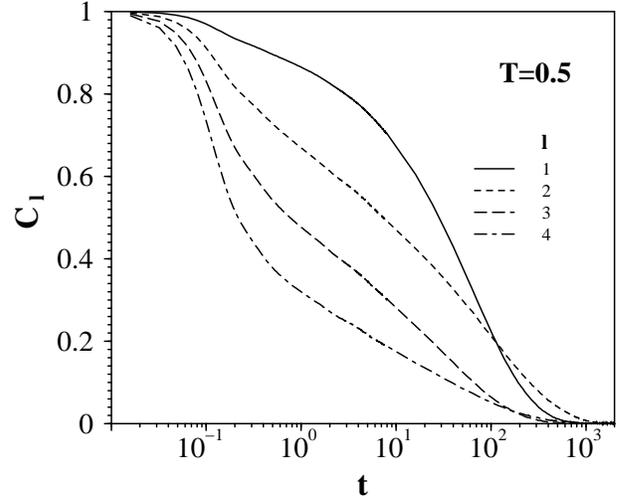,width=8cm,height=7cm}
\caption{Correlation functions $C_1$,$C_2$, $C_3$ and
$C_4$ at $T=0.5$. $C_l$  decay more at short times by increasing $l$. 
The curves with $l=1,3$ decay faster at long times than the curves 
with $l=2,4$, respectively, due to the partial head-tail symmetry of 
the molecules.}
\label{fig1234}
\end{figure}

Fig. \ref{fig1234} compares the four correlation functions $C_l$ 
with $l=1-4$ at $T=0.5$ are shown. Both the larger correlation 
loss at short times at larger $l$ values and the odd-even effect on the 
long-time decay of the correlations are evidenced.

\subsubsection{Angular velocity}

For a linear molecule the angular velocity is:

\begin{equation}
\omega = {\bf u} \times \dot{{\bf u}}
\label{omega}
\end{equation}

\noindent A set of correlation functions is defined as:

\begin{equation}
\Psi_l(t) = \frac{1}{N}\sum_{i=1}^{N}\langle P_l(\cos\alpha_i(t))
   \rangle,\quad l\geq1,  
\end{equation}

\begin{figure}
\psfig{file=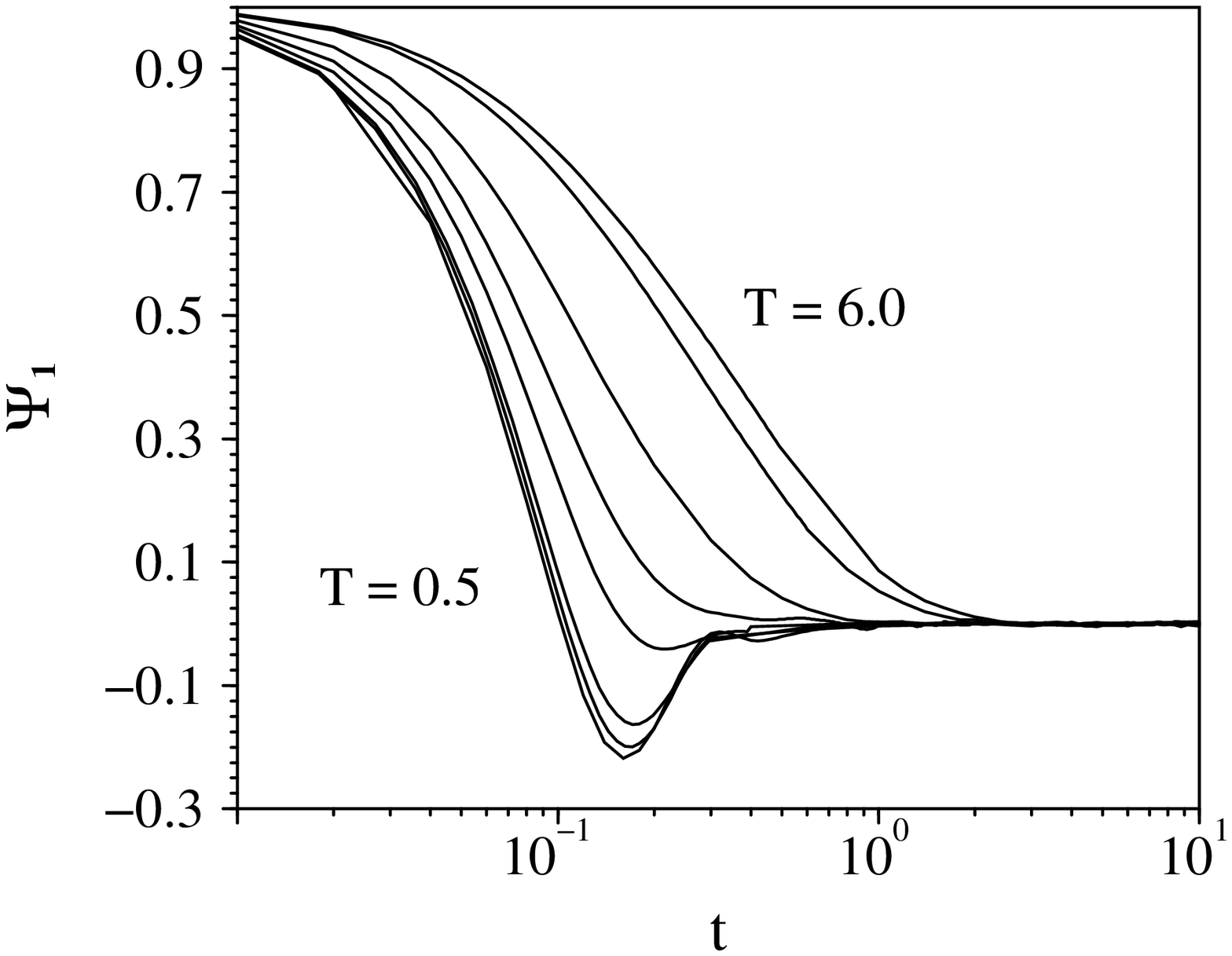,width=8cm,height=6.5cm}
\psfig{file=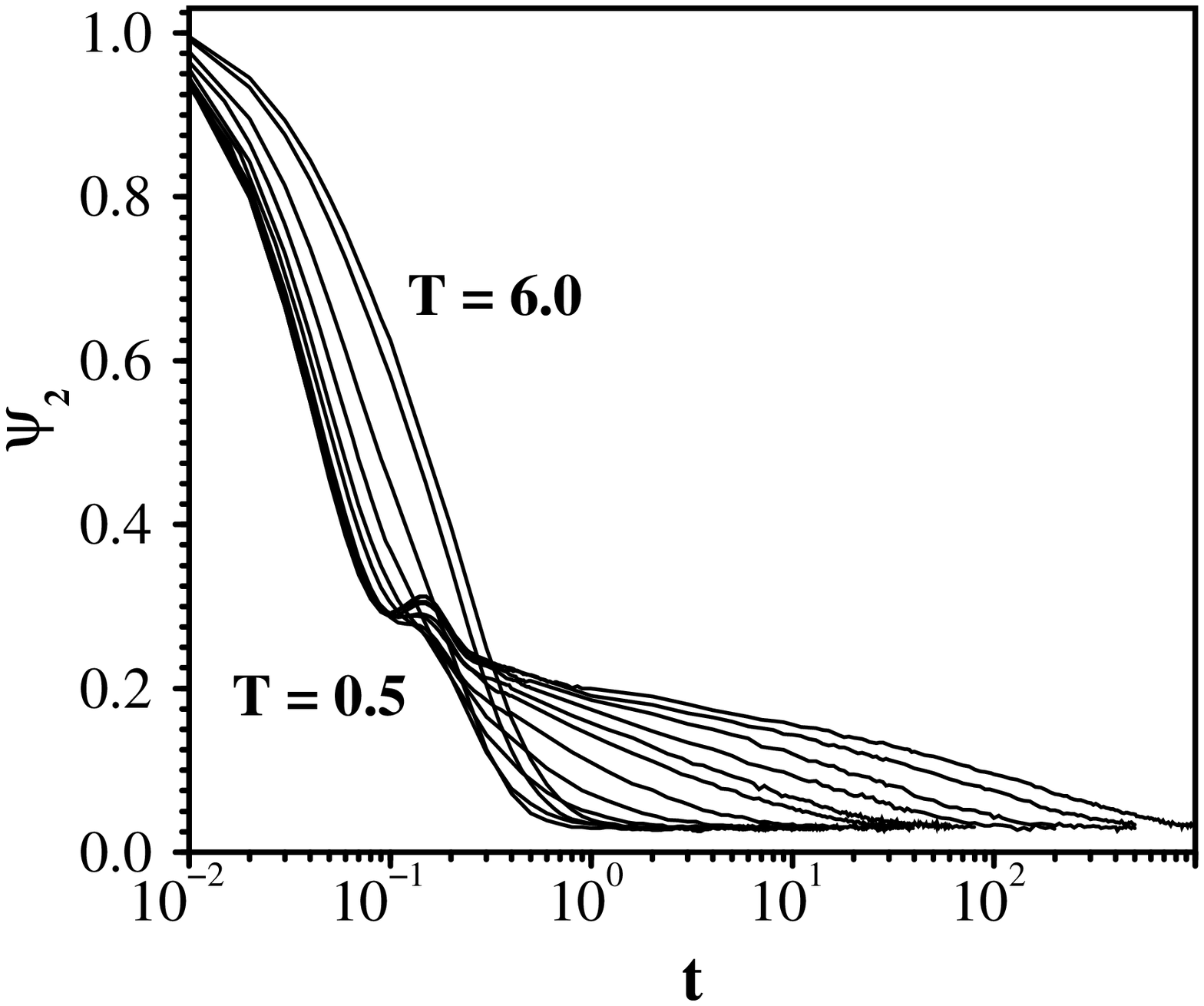,width=8cm,height=6.5cm}
\caption{Correlation functions of the angular velocity
$\Psi_1$ and $\Psi_2$ plotted for all the temperatures studied.}
\label{psi}
\end{figure}

\noindent $\alpha_i(t)$ is the angle between ${\bf \omega}_i(t)$ and
${\bf \omega}_i(0)$. In particular, for $l=1$ one has:

\begin{equation}
\Psi_1 = \frac{\meanm{{\bf\omega}(0)\cdot{\bf\omega}(t)}}
	{\meanm{|{\bf\omega}|^{2}}}
\end{equation}

\noindent which is the usual correlation function of the angular velocity.
In fig. \ref{psi} $\Psi_1$ and $\Psi_2$ are drawn
for all temperatures investigated. $\Psi_1$ decays fast and increasing $T$
slows down the decay  ( note the difference with the orientation 
case ). In particular, in the free-rotator limit $\Psi_1$ is a 
constant. At lower temperatures $\Psi_1$ shows a negative part at 
short times which evidences a change of sign of $\omega$. This must be
ascribed to the collisions on the rigid cage trapping the molecule 
( see figs. \ref{figc12} and \ref{figc34} ). An analogous effect was 
also noted for the linear velocity correlation function in I.
More insight on the rotational trapping may be gained by inspecting 
$\Psi_2$ in fig.\ref{psi}. No significant differences between $\Psi_2$  
and $\Psi_1$ are seen at high temperature. At lower 
temperatures, after a similar ballistic initial decay of 
$\Psi_1$ and $\Psi_2$, the latter slows down when 
$\Psi_2 \approx 0.25$. The long-living tail which shows up is  
interpreted by noting that at lower temperatures, after 
the ballistic regime the angular velocity is approximately trapped in 
a circle ( see eq.\ref{omega} ). 
An elementary calculation shows that $\Psi_1(t) = 0$ whereas $\Psi_2(t)
= 0.25$ as long as the trapping is effective \cite{renner}. When the 
molecular rearrangement allows the orientation relaxation, the angular 
velocity tends to be distributed over a sphere and $\Psi_2$ vanishes 
approximately as $C_{2}$.

\subsection{Diffusion coefficient and relaxation times}
\label{sec:diffrelax} 

The rotational diffusion coefficient of a linear molecule 
may be defined by a suitable Green-Kubo formula in close analogy 
to the translational counterpart as \cite{hansen}:

\begin{equation}
D_{r} = \frac{1}{2} \int_{0}^{\infty} 
\meanm{{\bf\omega}(0)\cdot{\bf\omega}(t)}dt
\end{equation}

From a computational point of view the evaluation of the above 
integral is delicate and it is more convenient to 
evaluate $D_{r}$ via the Einstein relation 

\begin{equation}
 D_r = \lim_{t\to\infty} \frac{R_{r}}{4t}
 \label{diffrotdef}
\end{equation}

\noindent $R_{r}$ is the  mean squared 
angular displacement :

\begin{equation}
R_r(t) = \frac{1}{N} \sum_{i=1}^{N}
       \langle \vert {\bf \phi}_i(t+t_0) - {\bf \phi}_i(t_0)\vert^2
       \rangle
\end{equation}

\noindent where ${\bf \phi}_i(t)$ is :

\begin{equation}
{\bf\phi}_i(t) - {\bf\phi}_i(0) = \Delta{\bf\phi}_i(t) = 
\int_{0}^{t} {\bf\omega}_i(t')\, dt'
\end{equation}

\begin{figure}
\psfig{file=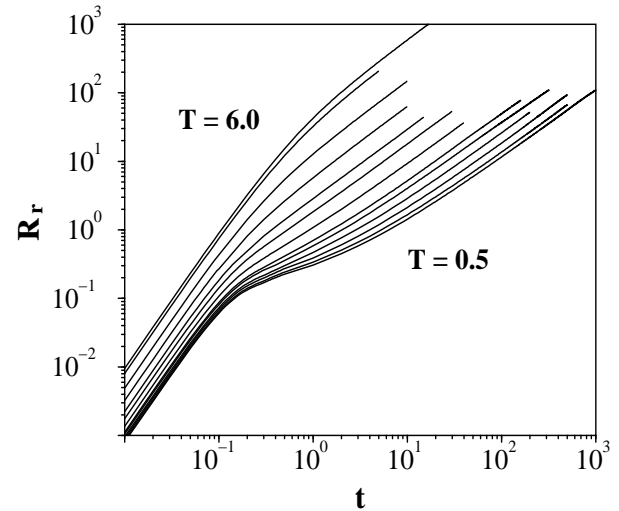,width=8cm,height=7cm}
\caption{ Mean squared angular displacement for all 
the temperatures investigated.}
\label{Rrot}
\end{figure}

\noindent In Fig. \ref{Rrot} $R_r(t) $ is shown.
The plots are qualitatively similar to the mean squared translational 
dislacement ( see I ). At short time the motion is ballistic. 
At intermediate times and lower temperatures a plateau shows up. It 
signals the increasing trapping of the molecular orientation due to 
the severe contraints on the structure relaxation. At longer times
the reorientation is diffusive according to eq.\ref{diffrotdef}.
By comparing $R_r(t) $ with the translational mean square 
displacement ( see fig.3 of I) it is seen that 
the angular trapping is weaker than the one affecting the 
center-of-mass motion since the subdiffusive intermediate regime 
is less pronounced and extends less on the time scale. 

The rotational correlation times are defined as \cite{hansen}:

\begin{equation}
\tau_{l} = \int_{0}^{\infty} C_{l}(t) dt 
\label{taudef}
\end{equation}

\begin{figure}
\psfig{file=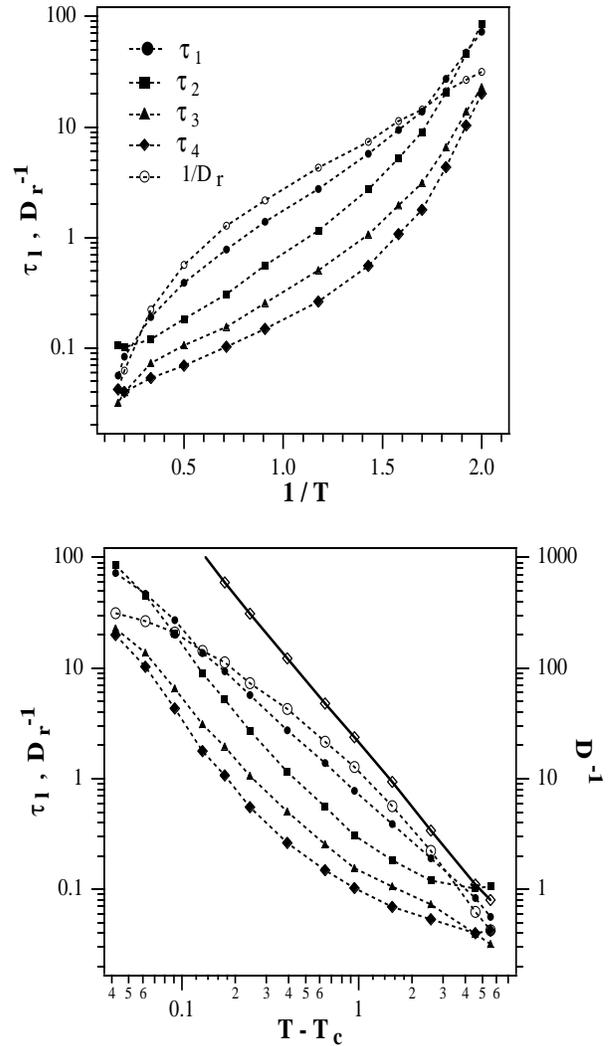,width=8cm,height=14cm}
\caption{Arrhenius plot ( top ) and MCT scaling analysis ( bottom ) 
of the rotational correlation times $\tau_{l}$, $l=1-4$ and the 
rotational diffusion coefficient $D_{r}$. $T_c = 0.458$. 
The dashed lines are guides for the eyes. The translational diffusion 
constant $D$ is also drawn for comparison ( open diamonds ). The 
continuous line is the best fit by using eq.11 with
$\gamma_{D} = 1.93 \pm 0.02$.}
\label{tauvsT}
\end{figure}

\noindent Fig. \ref{tauvsT} presents the T-dependence of $\tau_{l}$, 
$l=1-4$ and $D_{r}$. It is seen that a wide region exists where the 
above quantities exhibit approximately the same Arrhenius behavior
( about $0.7 < T < 2$ ). At lower temperatures the apparent activation 
energy of the rotational correlation times increase. In particular, 
as  noted in Sec.\ref{sec:orient}, $\tau_{1}$ becomes shorter than 
$\tau_{2}$ and a similar crossover is anticipated between $\tau_{3}$ 
and $\tau_{4}$ at temperatures just below $0.5$. Differently, the 
rotational diffusion coefficient $D_{r}$ exhibits the same activated 
behavior over a region which was shown to extend also below the 
critical temperature $T_{c}$ predicted by the mode-coupling theory 
( MCT ) \cite{kammererrot1}. 
The decoupling of $D_{r}$ with respect to $\tau_{l}$ may be 
anticipated by noting that the former is related to the area below 
$\Psi_1(t)$ ( eq.\ref{diffrotdef} ) and the latter to the area below 
$C_{l}(t)$ ( eq.\ref{taudef} ). At lower temperatures $\Psi_1(t)$ vanishes 
faster so probing the fast dynamics of the supercooled liquid whereas 
the decay of $C_{l}(t)$ slows down more and more ( see Sec. 
\ref{sec:corrfunc} ). Alternatively, it has to be noted that even in 
highly-constrained liquids small angular motions which are unable to 
relax the orientation lead to a finite value of $D_{r}$ in view 
of eq. \ref{diffrotdef} \cite{kammererrot1}. Such 
librational motions were detected in an MD study of OTP \cite{lewis}.

Fig. \ref{tauvsT} shows also the MCT analysis of the T-dependence of 
the correlation time and the rotational diffusion
\cite{schillingrot,gotzerot}. According to MCT, both $\tau_{l}$ 
and $D_{r}$ should scale as

\begin{equation}
\tau_{l}, D_{r}^{-1} \propto (T - T_{c})^{-\gamma}
\label{tauMCT}
\end{equation}

\begin{figure}
\psfig{file=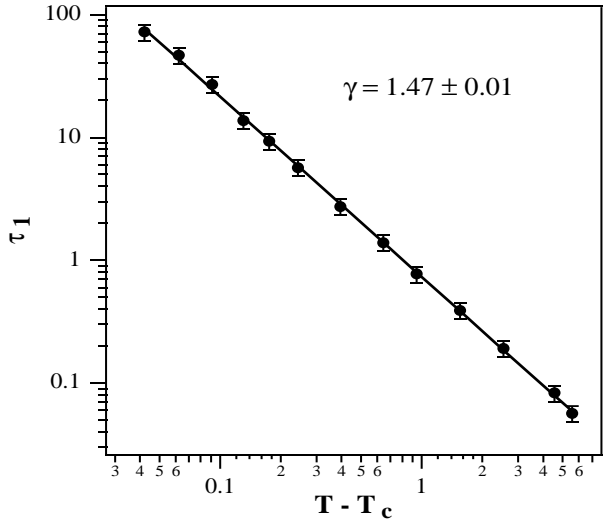,width=8cm,height=7cm}
\caption{MCT scaling analysis of $\tau_1$. $T_c = 0.458$.}
\label{tau1vsT}
\end{figure}

\noindent The underlying expectation on the scaling \ref{tauMCT} is that it
should work with the same $T_{c}$ value for any transport coefficient and
relaxation time. Differently, the physical meaning of $T_{c}$ could be 
weakened. In I it was shown that eq.\ref{tauMCT} fits the divergence 
of the translational diffusion coefficient $D$ over four orders of magnitude 
with $T_c = 0.458 \pm 0.002$ and $\gamma_{D} = 1.93 \pm 0.02$ ( the data 
are partially shown in fig. \ref{tauvsT} ) and that at lower temperatures
the primary relaxation time $\tau_{\alpha} \propto D^{-1}$. 
Fig. \ref{tauvsT} shows that the scaling \ref{tauMCT} 
is also effective for $\tau_{1}$  with 
$\gamma = 1.47 \pm 0.01$ ( see also fig.\ref{tau1vsT} ). However, 
meaningful deviations are apparent for $l \geq 2$ and $D_{r}$. The
deviations increase as $l$ increases. Since the rotational correlation 
functions $C_{l}$ are more and more sensitive to small-amplitude 
reorientations and then to small molecular displacements, the poorer 
scaling may be a consequence of the difficulties of the 
mode-coupling theories at short distances \cite{gotzevisco,balucani}.

\begin{figure}
\psfig{file=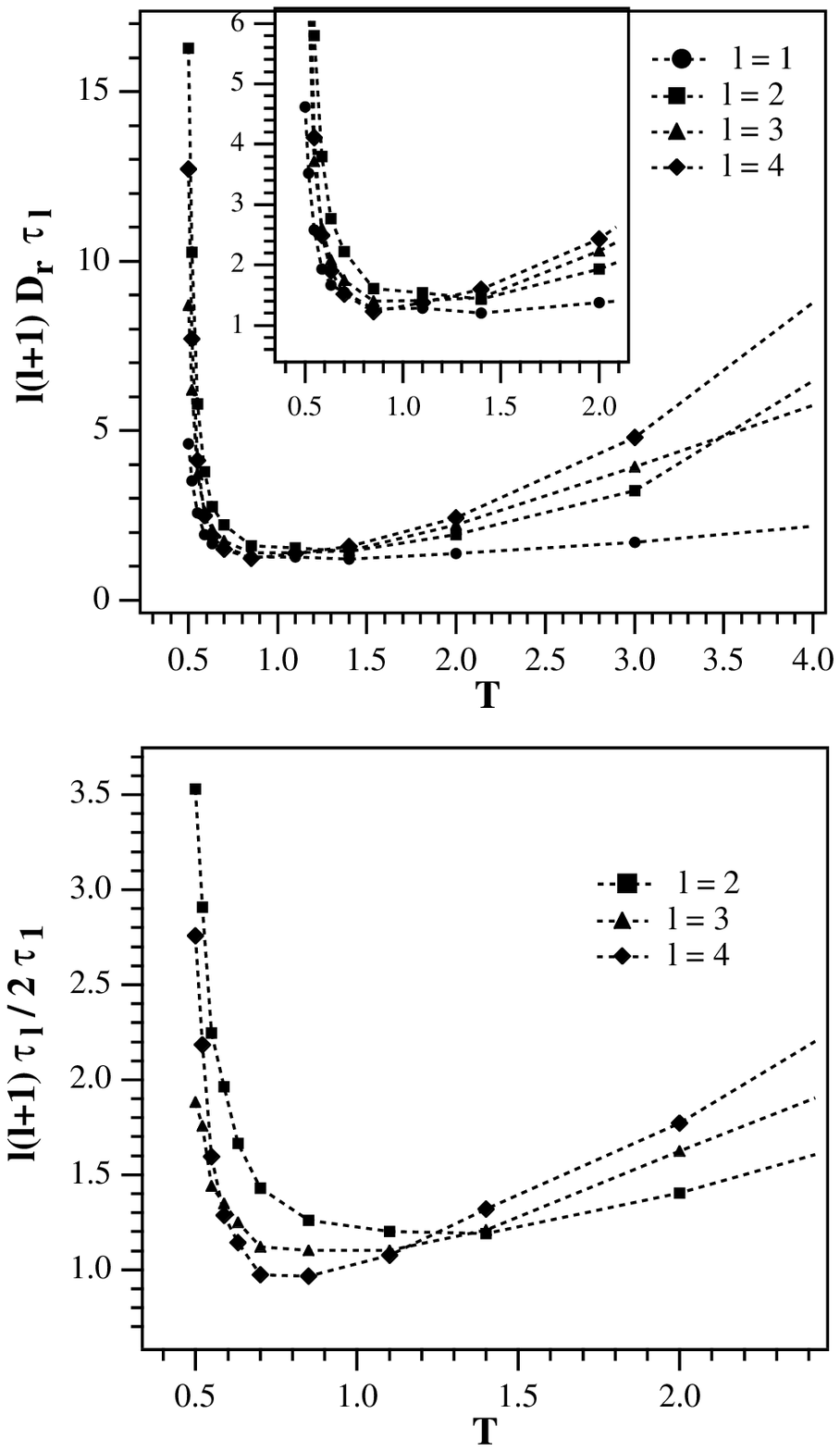,width=8cm,height=15cm}
\caption{Temperature dependence of the quantity $l(l+1) D_r \tau_l$
for  $l=1,2,3,4$ (top) and the ratio $l(l+1) \tau_l / 2 \tau_1$
for  $l=2,3,4$ (bottom). If the reorientation is diffusive both 
quantities must be equal to $1$. If the 
reorientation is jump-like $\tau_l \approx \tau_1  $.}
\label{checkdiff}
\end{figure}

To characterize the reorientation process, we studied the quantity 
$l(l+1) D_r \tau_l$ and the ratio $l(l+1) \tau_l / 2 \tau_1$. If the 
the molecule rotates by small angular jumps, or equivalently the waiting 
time in a single angular site is fairly shorter than the 
correlation time $\tau_{l}$, the motion is said to be diffusive and
both quantities are equal to $1$ for any $l$ value
\cite{favro,hansen}. The results are 
shown in fig.\ref{checkdiff}. Three regions may be broadly 
defined. For $T > 2$ the properties are gas-like and the 
rotational correlation times become fairly long.
For $0.7 < T < 2$ $l(l+1) D_r \tau_l$ and $l(l+1) \tau_l / 2 \tau_1$
do not change appreciably and are in the range $1-2$. For $T<0.7$
the above quantities diverge abruptly. The rapid increase in the deeply
supercooled regime demonstrates the failure of the diffusion model 
which in fact is expected to work only in liquids with moderate 
viscosity or if the reorientating molecule is quite large. If the 
assumption of small angular jumps is released and proper account of 
finite jumps with a single average waiting time in each angular site 
is made, the socalled jump-rotation model is derived \cite{ivanov}.
The main conclusion is that $\tau_{l}$ is roughly independent of 
$l$. In fact, for $l=2$ it is found that the quantity $l(l+1) 
\tau_l / 2 \tau_1 \approx 3.5$ at $T=0.5$ ( see fig.\ref{checkdiff} )
and the jump-rotation model predicts a value of about $3$. 
However, at higher $l$ values the comparison becomes much less 
favourable. For $l=3$ $l(l+1) \tau_l / 2 \tau_1 \approx 1.9$ at $T=0.5$
whereas the prediction is about $6$. For $l=4$ $l(l+1) \tau_l / 2 
\tau_1 \approx 2.75$ to be compared to the prediction is about $10$.
The failure of the usual simple rotational models is 
not unexpected. Their basic assumptions are rather questionable in 
supercooled liquids, e.g. the inherent homogeneity of the liquid and 
the presence of a single time scale both leading to the simple 
exponential decay of the rotational correlation functions.

\subsection{Jump rotation}

The inadequate description provided by the diffusion and the jump 
model calls for a further characterization of the rotational motion. To 
this aim we consider the self-part of the angular Van Hove function:

\begin{equation}
G_s^\theta(\theta,t) = 
\frac{2}{N\sin\theta}\sum_{i=1}^N \delta(\theta - \theta_i(t))
\end{equation}

\noindent $\theta_{i}(t)$ is the angle between the molecular axis of the 
i-th molecule at the initial time and time $t$ . 
$1/2 G_s^\theta(\theta,t) sin \theta d\theta$ is the probability to have 
the axis of a molecule at angle between $\theta$ and $\theta + d\theta$ 
at time $t$ with respect to the initial orientation. At long times 
$G_s^\theta(\theta,t) \cong 1$ since all the orientations are equiprobable.

\begin{figure}
\psfig{file=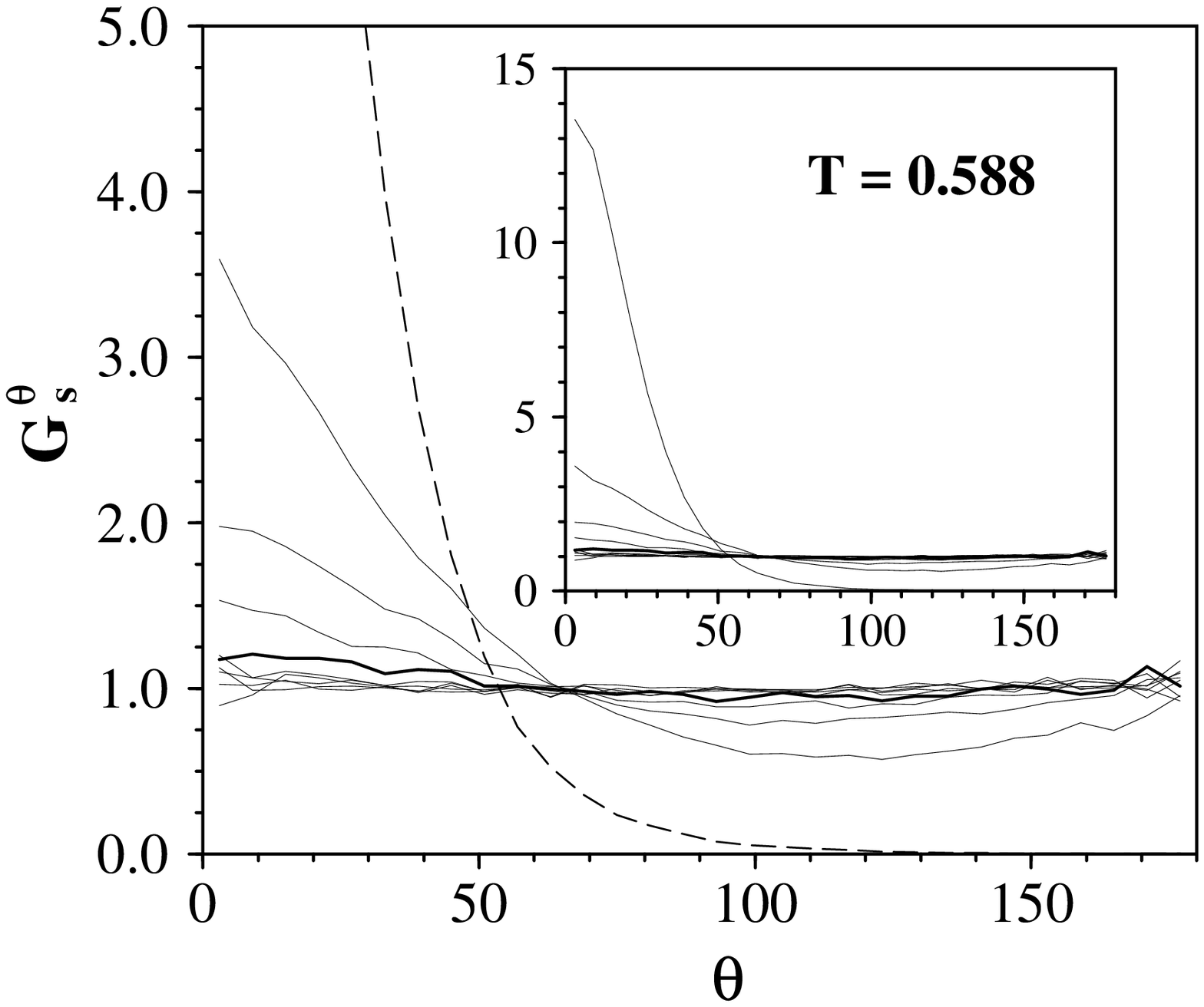,width=8cm,height=6.5cm}
\psfig{file=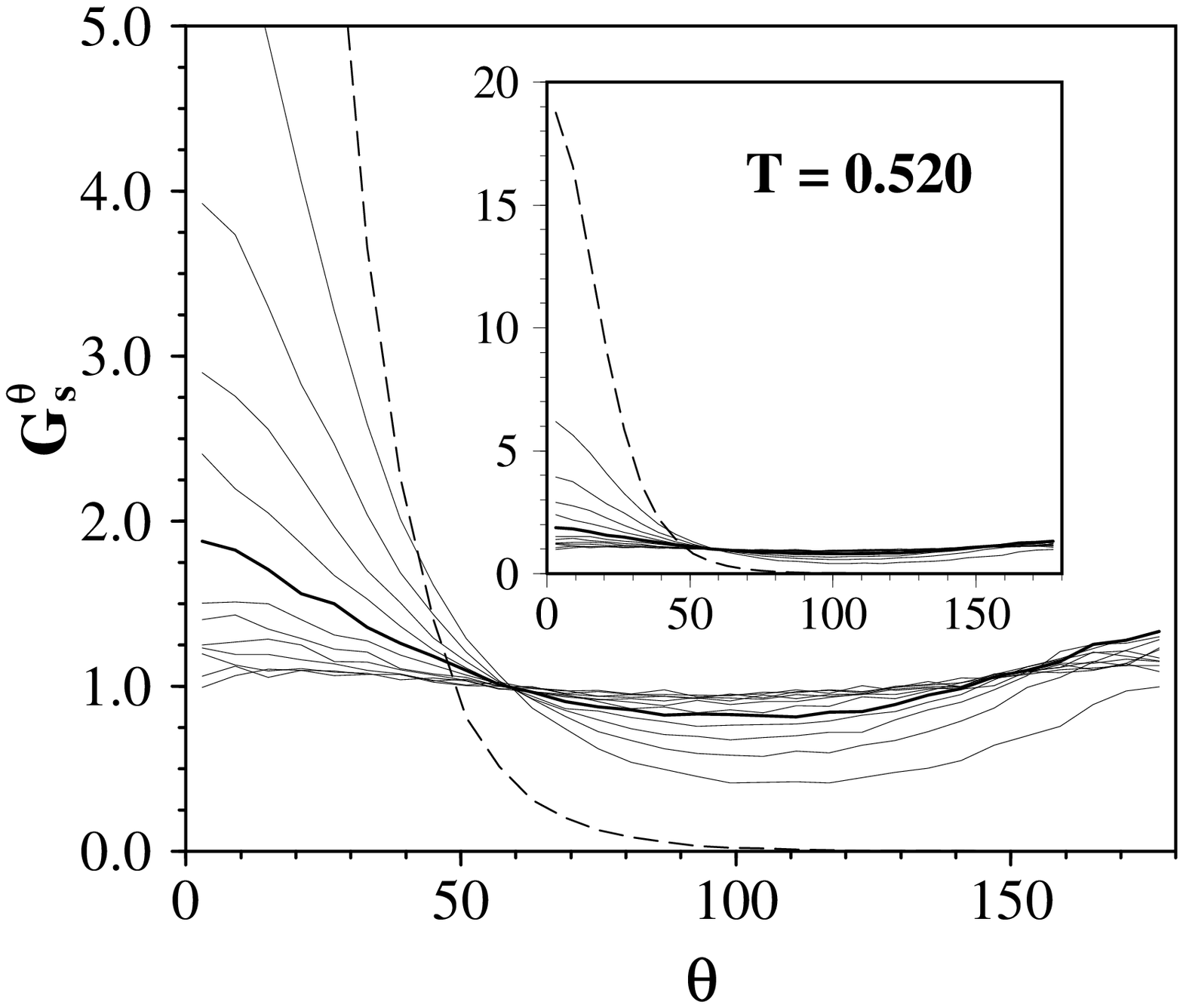,width=8cm,height=6.5cm}
\psfig{file=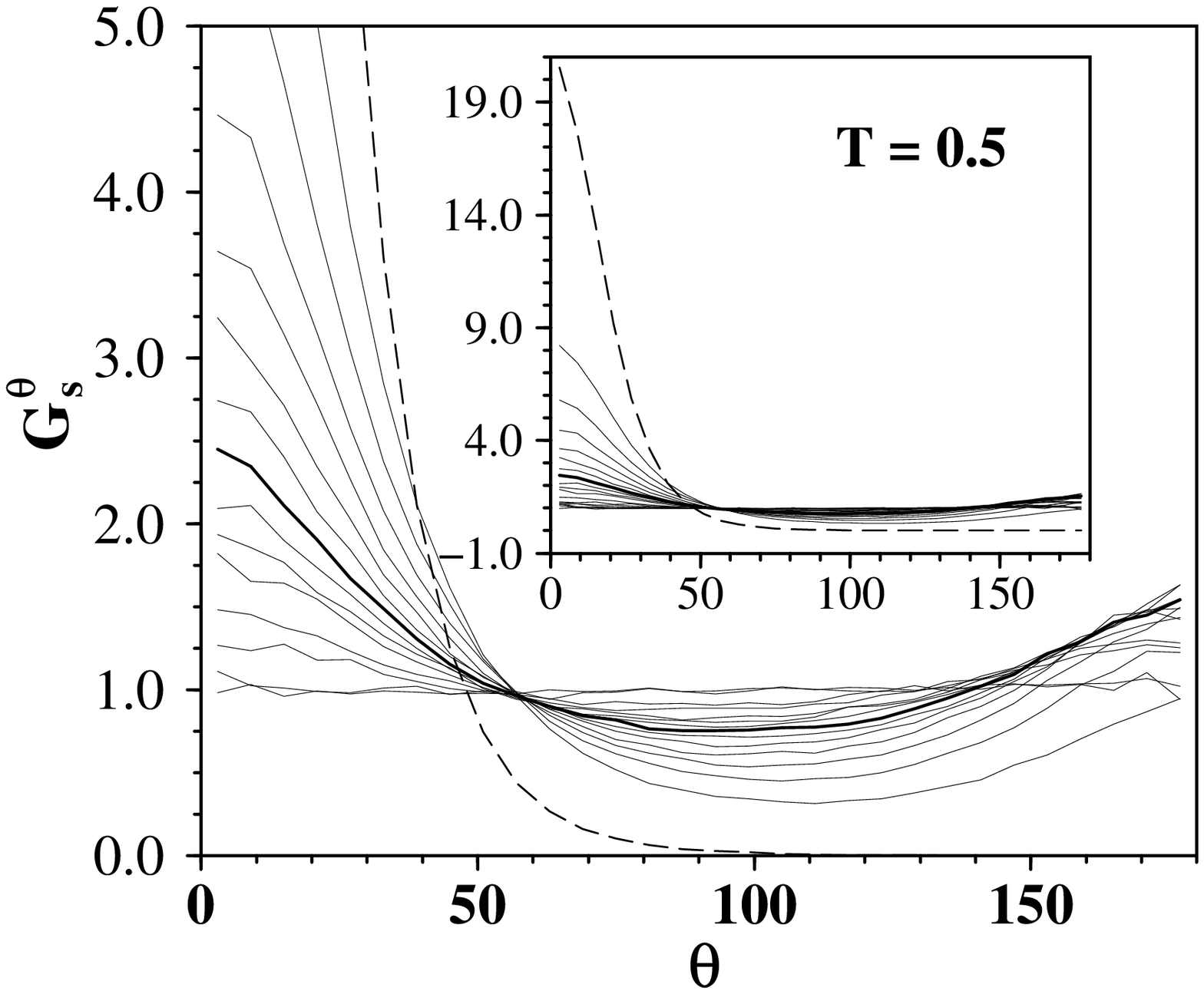,width=8cm,height=6.5cm}
\caption{The angular Van Hove functions at low temperatures.
Top: $T=0.588$, bold line $t=80$.
Middle: $T=0.520$, bold line $t=200$.
Bottom: $T=0.5$, bold line  $t=280$. Dashed line $t=1$. The other 
continuous lines are plotted at intermediate times with equal spacing.}
\label{vanhove}
\end{figure}

In Fig. \ref{vanhove} the function $G_s^\theta$ is plotted for
different temperatures and several times. At higher
temperatures , as the time goes by, the molecule explores more and 
more angular sites in a continuous way. Instead, at lower temperatures 
$G_s^\theta$ exhibits a peak at 
$\theta \approx 180^{\circ}$ and intermediate times signaling that the 
reorientation has a meaningful probability to occur by jumps.

\begin{figure}
\psfig{file=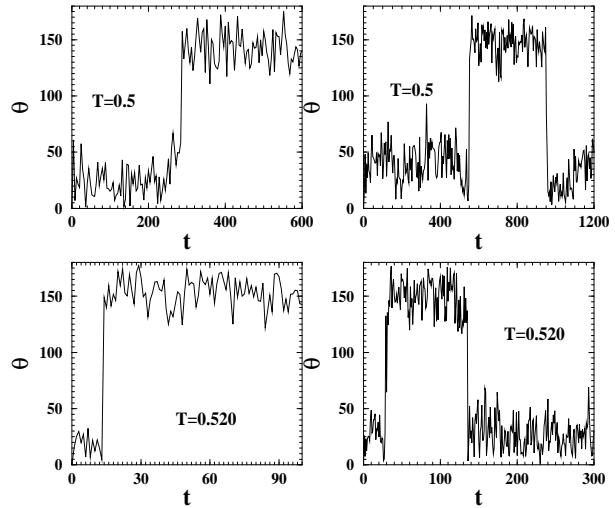,width=8cm,height=7cm}
\caption{Selected single-particle orientations evidencing the flips 
occurring at low temperatures.}
\label{jumps}
\end{figure}

The indications provided by the VanHove function concerning the 
presence of rotational jumps are confirmed by directly inspecting 
the single particles trajectories ( fig. \ref{jumps} ). Similar 
findings were reported also in other studies on dumbbells and CKN 
glassformers \cite{kammererrot1,signorini}. With respect to 
the translational counterparts, it must be pointed out that they are 
quite faster ( see I ) and more frequent ( about one order of 
magnitude). The higher number of rotational jumps is also 
anticipated by noting that, differently from the translational Van-Hove 
function, the rotational one does exhibit explicit signatures of jump 
motion ( see I ).

To characterize the jumps we studied the distribution $\psi_{rot}(t)$
of the waiting-time, namely the residence time in one angular site
of the unit vector ${\bf u}_{i}$ being parallel to the axis of the 
i-th molecule. 
A jump of the i-th molecule is detected at $t_{0}$ if the angle between 
${\bf u}_{i}(t_{0})$ and ${\bf u}_{i}(t_{0} +\Delta t^{*} )$ is larger 
than $100^{\circ}$ with $\Delta t^{*} = 24$. To prevent
multiple countings of the same jump, the molecule which jumped at time $t$ 
is forgotten for a lapse of time $\Delta t ^{*}$. To minimize possible 
contributions due to fast rattling motion, each angular displacement
is averaged with the previous and the next ones being spaced typically by 
$6-8$ time units, depending on the temperature. The jump search procedure 
was validated by inspecting several single-molecule trajectories. 
The above definition of rotational jump fits well 
their general features, i.e. they are rather 
fast and exhibit no meaningful distribution of both the amplitude 
and the time needed to complete a 
jump ( see fig.\ref{jumps} ). It is worth noting that 
in I it was found that the time needed to complete the translational 
jumps exhibits a distribution. The absence of a similar distribution for 
the rotational jumps points to a larger freedom of the latter.

\begin{figure}
\psfig{file=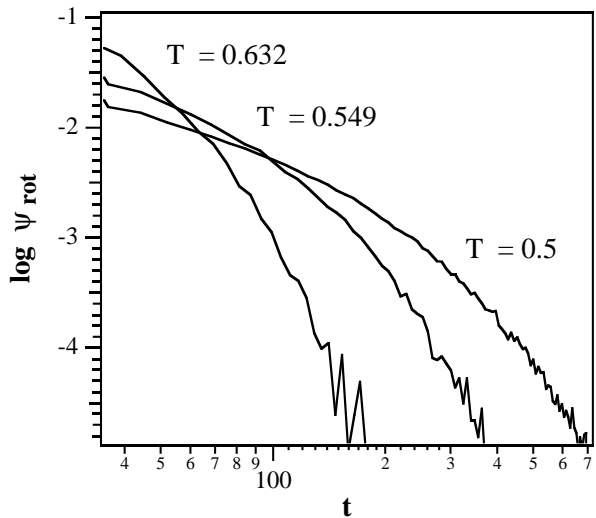,width=8cm,height=7cm}
\caption{The rotational waiting-time distribution $\psi_{rot}(t)$ 
at low temperatures.}
\label{wtd}
\end{figure}

\begin{figure}
\psfig{file=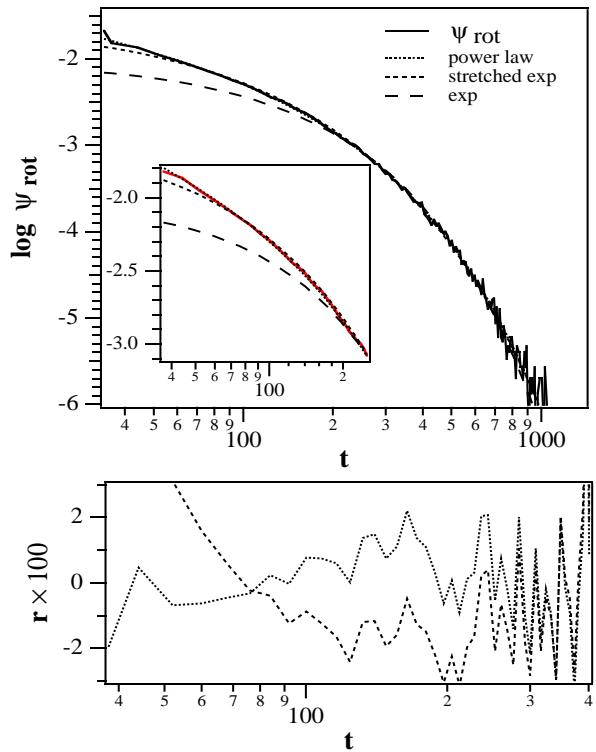,width=8cm,height=10cm}
\caption{Analysis of the rotational waiting-time distribution 
$\psi_{rot}$ at $T=0.5$. Top: comparison of the best fits with 
the exponential ( $\tau = 102 \pm 3$ ), the 
stretched exponential ( stretching $\beta = 0.78 \pm 0.02$ and
$\tau = 63 \pm 4$ ) and the truncated power law eq.13 (
$\xi = 0.34 \pm 0.04$ and $\tau = 125 \pm 3$ ). The insert is a 
magnification of the short-time region. Notice that the fit with the 
power law is virtually superimposed to $\psi_{rot}$.
Bottom: residuals of the fits in terms of
the truncated power law and the stretched exponential.}
\label{wtd05}
\end{figure}

Fig.\ref{wtd} shows $\psi_{rot}(t)$ at different temperatures. At 
$T=0.632$ is virtually exponential. At lower temperatures deviations 
become apparent which are analyzed for $T=0.5$ in fig. \ref{wtd05}.
In I it was noted that the translational waiting-time 
distribution may be fitted nicely by the truncated power law:

 \begin{equation}
\psi(t) =  \left[\Gamma (\xi )\tau^{\xi}\right]^{-1}
t^{\xi-1} e^{ -t / \tau } \hspace{1cm} 0 < \xi \le 1
\label{power}
\end{equation}
 
\noindent The best fit provided by eq.\ref{power} is compared to the fits 
by using the stretched ( $exp[-(t/\tau)^{\beta}]$ ) and the usual 
exponential functions in fig. \ref{wtd05}. The better agreement of 
eq.\ref{power} at short times may be appreciated by looking at the 
residuals.

The fractal behavior of both the translational 
and the rotational waiting-time distributions is an
indication that the molecular motion at short times exhibits 
intermittent behavior. The issue in the framework of glasses 
has been addressed by several 
authors \cite{sjogren,hubbard,odagaki,sharon,muranaka}. 
The exponent $\xi$ of eq.\ref{power}
has a simple interpretation. If a dot on the time axis 
marks a jump, the fractal dimension 
of the set of dots is $\xi$. For $\xi < 1$, it follows
$\psi(t) \propto t^{\xi-1}$ at short times \cite{sjogren,hilfer} in 
agreement with our results. If $\xi =1$, the distribution 
of dots is uniform and $\psi(t)$ recovers the exponential form. 
This is expected beyond a time scale $\tau$ and 
the exponential decay of $\psi_{rot}(t)$ at long times signals the 
crossover to the usual Poisson regime. It must be noted that 
the translational waiting time distribution at the lowest 
temperature ( $T=0.5$ ) shows a weak 
tendency to vanish faster than eq.\ref{power}. A similar feature is 
not observed in $\psi_{rot}(t)$.

\subsection{Breakdown of the Debye-Stokes-Einstein Law}

For large Brownian particles the reorientation in a liquid
occurs via a series of small angular steps, i.e. it is diffusive. 
Hydrodynamics predicts that the diffusion manifests a strong 
coupling to the viscosity $ \eta $ which is accounted for by the 
Debye-Stokes-Einstein law ( DSE ). For biaxial 
ellipsoids it takes the form \cite{favro}

\begin{equation}
D_i = \frac{kT}{{\mu}_i\eta}, \hspace{1cm} i=x,y,z
\label{eq:DSE}
\end{equation}

\noindent $D_{x,y,z}$ are the principal values of the diffusion 
tensor, $k$ is the Boltzmann constant. The coefficients $\mu_i$ depend on 
the geometry and the boundary conditions ( BC ). For a sphere with stick 
BC $\mu_{x,y,z} = 6v$, $v$ being the volume of the sphere. For an
uniaxial ellipsoid one considers $D_{\parallel} = D_{z}$ and 
$D_{\perp} = D_{x} = D_{y}$. The case of stick BC can be worked 
analtycally \cite{lamb,favro}. For slip BC numerical results for 
$D_{\perp}$ are known \cite{hu} ( note that in this case the fluid does 
not exert torques parallel to the symmetry axis ). 
Eq.\ref{eq:DSE} is sometimes rewritten in an alternative form in terms of 
proper rotational correlation times, e.g. for uniaxial 
molecules the equality $\tau_{l} = 1/l(l+1) D_{\perp}$ holds. 
The new form is more suitable for comparison with the experiments since
they do not usually provide direct access to the rotational diffusion 
coefficients.

Irrespective of the heavy hydrodynamic assumptions, DSE works nicely even 
at a molecular level if the viscosity is not high ( $\eta < 1 Poise$ ). 
Deviations are observed at higher viscosities for tracers in supercooled 
liquids by time-resolved fluorescence \cite{ye,tork2} and Electron Spin 
Resonance ( ESR ) studies \cite{euro,vigo,macromol}.  On the other hand, 
photobleaching \cite{ediger} and NMR \cite{sillescu1},
studies found only small deviations from DSE even close to $T_{g}$.
In all the cases known, DSE is found to overestimate the rotational 
correlation times since on cooling their increase is less than the one 
being exhibited by the viscosity. In this decoupling region ESR  
evidenced that the tracer under investigation rotates by jump motion 
\cite{jumpesr}. 

Shear motion facilitates molecular jumps over energy barriers \cite{eyring}. 
On the other hand, guest molecules may jump in frozen hosts in the 
absence of viscous flow. Since a meaningful fraction of the molecule 
in the system reorientates by finite angular steps with intermittent 
behavior, not quite expected in a liquid, it is of interest to investigate 
to what extent the reorientation is coupled to the viscous shear flow.  

\begin{figure}
\psfig{file=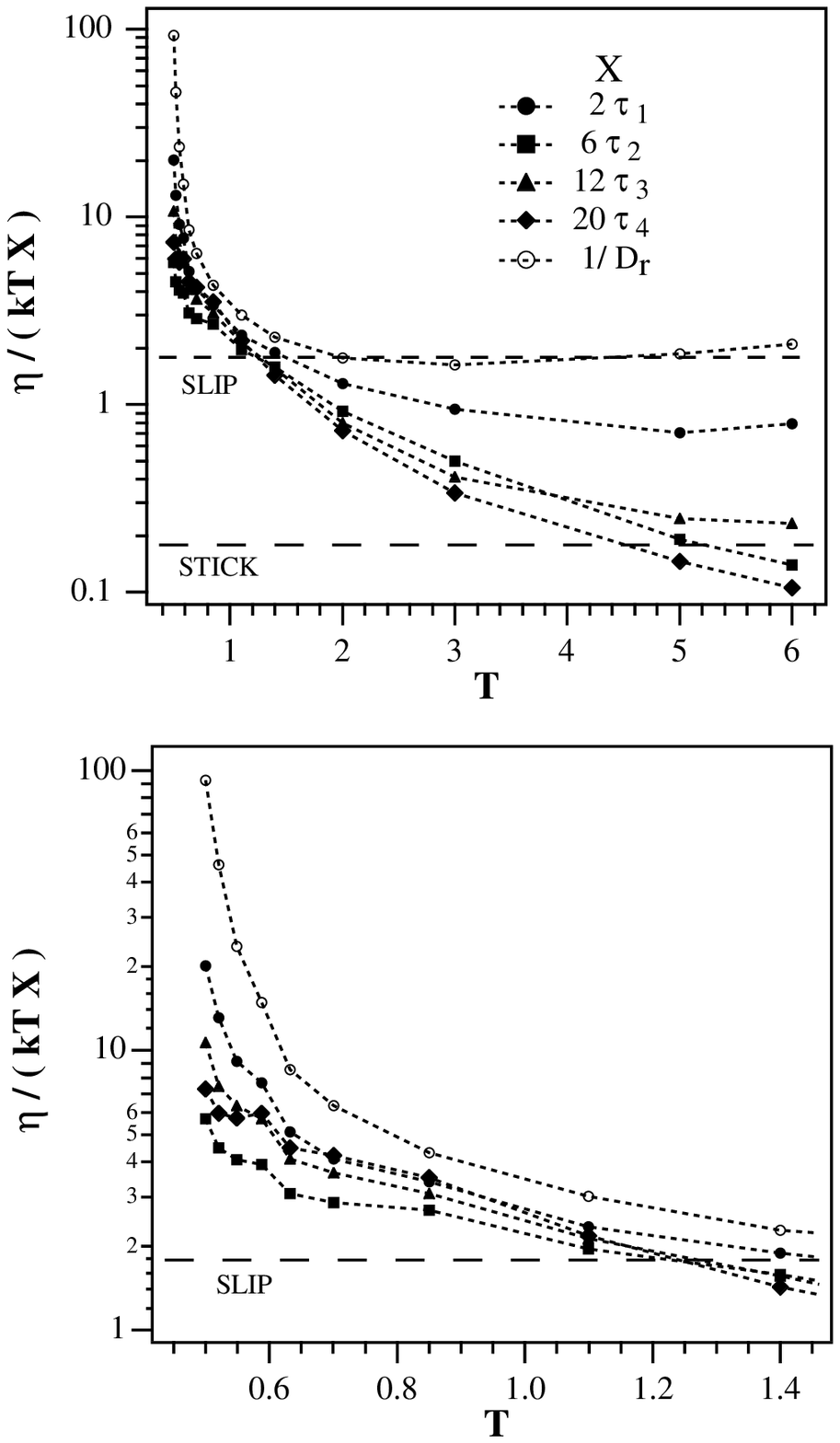,width=8cm,height=15cm}
\caption{ Plots of the quantity $\eta/X kT$ with $X = D_{r}^{-1}, 
l(l+1) \tau_{l}$ over the overall temperature range ( top ) and in 
the supercooled regime ( bottom ). According to DSE, the quantity is a 
constant whose value depends on the hydrodynamic boundary conditions. 
The superimposed dashed lines are the DSE expectations for 
stick and slip boundary conditions. Notice that at $T=1.4$ $\eta/X 
kT$ depends little on $X$, signaling diffusive behavior. }
\label{fig:DSE}
\end{figure}

The results are shown in fig. \ref{fig:DSE} by plotting the 
the quantity $\eta/X kT$ with $X = D_{r}^{-1}, l(l+1) \tau_{l}$
with $l=1-4$. According to DSE it should be constant. The 
viscosity data were taken from I. At high 
temperatures the quantity approaches the value expected 
for stick BC. For $T > 5$ a tendency of $\eta/X kT$ for 
$X=D_{r},\tau_{1}$ to increase is noted. However, at such 
temperatures the system manifests gas-like features ( see 
figs.\ref{figc12},\ref{figc34},\ref{psi} ) .
On cooling $\eta/X kT$ increases. For intermediate temperatures the 
liquid properties are well developed, the system is diffusive 
( $l(l+1)\tau_{l} D_{r} \approx 1$, see fig.\ref{checkdiff} ) and 
$\eta/X kT$ has a value close to the DSE 
expectation with slip BC. Notably, $D_{r}\eta/ kT$ remains close to
this value in the wide interval $2<T<6$. At lower temperatures  
$\eta/X kT$ diverges. The stronger deviations are exhibited by 
$D_{r}$ and $\tau_{1}$, the weaker ones by $\tau_{2}$. $\tau_{3}$ and 
$\tau_{4}$ track the behavior of $\tau_{1}$ and $\tau_{4}$, 
respectively, being the pair $\tau_{1,3}$ being affected by the jump 
motion much more than the pair $\tau_{2,4}$ ( see sec.\ref{sec:orient}. 
$\eta$ increases of a factor of about $400$ between $T=1.4$ and $0.5$. 
The corresponding changes of $D_{r}\eta/ kT$, $\eta/\tau_{1} kT$ 
and $\eta/\tau_{2} kT$ are $41$, $11$ and $3.6$, respectively.
The discussion support the conclusion that the correlators being 
affected by rotational jumps, e.g. $C_{1,3}(t)$, yield correlation 
times fairly more decoupled by the viscosity.

If one compares the changes of $\eta/\tau_{2} kT$ to the ones drawn by
ESR and fluorescence experiments in the region $T/T_{c} \approx 1.1-1.5$, 
a broad agreement is found \cite{ye,euro}. These experiments found 
even larger values of $\eta/\tau_{2} kT$ on approaching $T_{g}$. Instead, 
photobleaching studies detected changes of less than order of 
magnitude by changing $\eta$ over about $12$ orders of magnitude 
which are extremely smaller than the present ones \cite{ediger}.

Inspection of fig.\ref{fig:DSE} suggests a way to reconcile, at 
least partially, the ESR \cite{euro}, the photobleaching 
\cite{ediger} and NMR \cite{sillescu1} studies on the glass former 
o-terphenyl (OTP). Both photobleaching and NMR 
measure $\tau_{2}$ in a direct way. Instead, in the glass transition 
region the ESR lineshape 
depends in principle on several $\tau_{l}$ and a model 
is needed to relate them to each other and lead to $\tau_{2}$ 
\cite{jumpesr}. The model is adjusted by fitting 
the teoretical prediction with the highly structured ESR lineshape.
Fig.\ref{fig:DSE} shows that the decoupling, as expressed by 
$\eta/\tau_{l} kT$ increases with $l$. Since the weight of the 
$\tau_{l}$ to set the ESR lineshape is roughly comparable, one may 
anticipate that $\tau_{2}^{ESR}$ may be underestimated at some degree
{\it around } the glass transition. On the other hand, it must be 
pointed out that the rotational decoupling is observed up to $1.4 T_{g}$
in OTP where ESR yields $\tau_{2}$ in a model-independent way \cite{euro}.
Furthermore, the decoupling is evidenced also by fluorescence experiments
which provide $\tau_{2}$ in a model-independent way  \cite{ye}.

\section{CONCLUSIONS}
\label{sec:conclusion}

The present paper investigated the rotational dynamics of a 
supercooled molecular system. The study addressed several general 
features and focussed on the characterization of the jump dynamics 
and the degree of coupling with the viscosity.

The ensemble consists of rigid A-B 
dumbbells interacting via a L-J potential. All the properties 
were studied along the isobar $P=1.5$. The time
orientation correlation functions $C_{l}(t)$ exhibit at high temperatures 
gas-like features. In the supercooled regime, after a first initial 
decay, a plateau is observed which signals the trapping of the molecule 
due to the increased difficulty of the surroundings to rearrange themselves. 
The plateau level decreases by increasing the $l$ rank of the correlation 
function due to the larger sensitivity to small-angle 
librations. The long-time decay of $C_{l}(t)$ is fairly well 
decscribed by a stretched exponential. The stretching increases from 
$l=1$ ( $\beta = 0.7$ ) to $l=4$ ( $\beta = 0.47$ ) at $T=0.5$.
The influence on the decay time of $C_{l}$ ( and then 
on $\tau_{l}$ ) of the partial head-tail symmetry of the dumbbells
was noted.  

A set of angular-velocity correlation functions $\Psi_{l}$ was 
defined. $\Psi_{1}$ decays faster and faster on decreasing $T$. 
Differently, $\Psi_{2}$ develops a long-lasting tail which 
vanishes on the time scale of $C_{2}$. 

The temperature dependence of the rotational correlation times 
$\tau_{l}$ with $l=2-4$ and the rotational diffusion coefficient $D_{r}$ 
manifest deviations by the power-law scaling in $T-T_{c}$, being 
$T_{c}$ the MCT critical temperature. There were ascribed to the 
difficulties which mode-coupling theories meet at short length
scales  \cite{gotzevisco,balucani}. Remarkably, 
the scaling works nicely for $\tau_{1}$ over more than three orders 
of magnitude with an exponent $\gamma = 1.47 \pm 0.01$. 
This parallels the scaling which was noted for the translation 
diffusion coefficient and the primary relaxation time 
$\tau_{\alpha}$ in I.
For $0.7 < T < 2$ the quantities $l(l+1) D_r \tau_l$ and 
$l(l+1) \tau_l / 2 \tau_1$
do not change appreciably and are in the range $1-2$ in good 
agreement with the diffusion model which predicts that both of them 
equal to $1$ independently of the temperature. For $T<0.7$ the above 
quantities increase abruptly. The increase of the quantity
$l(l+1) \tau_l / 2 \tau_1$ is reasonably accounted for by the 
jump-rotation model for $l=2$ \cite{ivanov}. For higher $l$ values the 
agreement becomes quite poor. Being $\tau_1$ and $\tau_2$ measured
by most experiments \cite{ediger,sillescu1,tork3,euro,tork2,williams1} 
this finding may account for the attention that the jump model has 
attracted during the last years \cite{jumpesr,williams2,williams3}.

The analysis of the angular Van Hove function evidences that 
in this region a meaningful fraction of the sample reorientates by
jumps of about $180^{\circ}$. The flips are rather 
fast and exhibit no meaningful distribution of both the amplitude 
and the time needed to complete a jump. Differently, it was noted 
in I that translational jumps require different times to be performed.
The absence of a similar effect for the reorientations indicates a 
larger angular freedom. This is also apparent by the larger number of 
rotational jumps which are detected with respect to the translational 
ones. It is worth noting that the ease to jump 
does not lead to trivial relaxation properties, as signaled by 
the stretched decay of $C_{1,2,3,4}$.

We characterized the distribution 
$\psi_{rot}$ of the waiting-times in the angular sites. It 
vanishes exponentially at long times whereas at lower temperatures it
decays at short times as $t^{\xi-1}$ with  
$\xi = 0.34 \pm 0.04$ at $T=0.5$. 
Interestingly, the translational waiting-time distribution exhibits 
the same behavior ( see I ). The exponent for the translational case 
is $\xi = 0.49$. We ascribe the power-law to the intermittent 
features of the motion in glassy systems 
\cite{sjogren,hubbard,odagaki,sharon}. 

The intermittent jump reorientation is fairly different from the 
motion in a liquid. Then, a decoupling from the viscous flow and the 
subsequent breakdown of the Debye-Stokes-Einstein is anticipated. Our 
study confirms the breakdown and shows that the quantity 
$\eta/X kT$ with $X = D_{r}^{-1}, l(l+1) \tau_{l}$ and $l=1-4$ 
diverges below $T=1$. In particular, the correlators being 
affected by rotational jumps, e.g. $C_{1,3}(t)$, yield correlation 
times fairly more decoupled by the viscosity. A rather similar 
decoupling was found in I for the product $D \eta$, $D$ being 
the translational diffusion coefficient.

The decoupling of the molecular reorientation by the viscosity 
could be also anticipated by the observed ease to perform 
rotational jumps. The reduced tendency to freeze of 
the rotational degrees of freedom was pointed out by MD 
\cite{bagchi} and theoretical \cite{schillingrot} studies.
The former investigating the residual rotational relaxation in a 
random lattice with {\it quenched translations} and the latter 
predicting a hyerarchy for the glassy freezing, i.e. the rotational 
dynamics can never freeze {\it before} the translational dynamics. 
The decoupling of the rotational motion of 
guest molecules from the viscous flow has been experimentally seen by
by time-resolved fluorescence \cite{ye,tork2} and Electron Spin Resonance 
\cite{euro,vigo,macromol} while photobleaching and NMR
studies reported small deviations from DSE even close to $T_{g}$ 
\cite{ediger,sillescu1}. It is worth noting that the decoupling of 
the translational diffusion from the viscosity and related phenomena 
as the so-called rotation-translation paradox have been ascribed to a 
spatial distribution of 
mobility and relaxation properties, so called dynamical heterogeneities 
\cite{ediger,tork3,sti,opp,douglas,sillescu3,claudio}. Their role will 
be addressed in a forthcoming study.

\acknowledgments

The authors warmly thank Walter Kob for having suggested the 
investigation of the present model system and the careful reading of 
the manuscript. Umberto Balucani , Claudio Donati and 
Francesco Sciortino are thanked for many helpful discussions and
Jack Douglas for a preprint of ref. \cite{sharon}.

\end{document}